\documentclass[twocolumn,twoside]{article}
\usepackage{Arxiv,multicol,times}

\usepackage{epsfig}
\usepackage{graphicx}
\usepackage{hyperref}
\usepackage{amssymb}
\usepackage{algorithm}
\usepackage{algorithmic}
\usepackage{inputenc}
\usepackage{amssymb}
\usepackage{amsmath}
\usepackage[overload]{empheq}
\usepackage{hyphenat}
\usepackage{float}
\usepackage{placeins}
\usepackage{multirow}

\usepackage[pagewise]{lineno}
\emergencystretch=\maxdimen
\PUBYEAR{2022}
\VOL{1}
\newcommand\startpage{1}
\usepackage{fancyhdr}
\usepackage{ifthen}
\usepackage{lastpage}
\fancypagestyle{mypagestyle}{%
    \fancyhf{} % clear all fields
    \fancyhead[RO]{Omar BOUDRAA}
    \fancyhead[LO]{\thepage}
    \fancyhead[LE]{Segmentation of 3D Dental Images Using Deep Learning}
    \fancyhead[RE]{\thepage }
    \renewcommand{\headrulewidth}{0pt}
}
\begin{document}
\fancypagestyle{plain}{%
  \renewcommand{\headrulewidth}{0pt}%
  \fancyhf{}%
  \fancyfoot[R]{1%Dynamic Publishers, Inc., USA%
  }
}
\sloppy

\title{
%\begin{small}Submitted: 22 Jan, 2019; Accepted: 1 Mar, 2019; Publish: XXX\end{small}\\
Segmentation of 3D Dental Images Using Deep Learning}

\author{{\bf Omar BOUDRAA$^1$} \\[1em]
	$^1$Department of Computer Science, XLIM, UMR CNRS 7252, University of Limoges,\\ 123 Avenue Albert Thomas, 87060 Limoges Cedex, France.\\
	\textit{omar.boudraa@gmail.com} }

\maketitle

\begin{abstract} 
	3D image segmentation is a recent and crucial step in many medical analysis and recognition schemes. In fact, it represents a relevant research subject and a fundamental challenge due to its importance and influence. This paper provides a multi-phase Deep Learning-based system that hybridizes various efficient methods in order to get the best 3D segmentation output. First, to reduce the amount of data and accelerate the processing time, the application of Decimate compression technique is suggested and justified. We then use a CNN model to segment dental images into fifteen separated classes. In the end, a special KNN-based transformation is applied for the purpose of removing isolated meshes and of correcting dental forms. Experimentations demonstrate the precision and the robustness of the selected framework applied to 3D dental images within a private clinical benchmark.
\end{abstract}


\section{Introduction}
\label{intro}
Segmentation is one of the critical steps in image analysis. It makes it possible to isolate in the image the objects to be analyzed and to separate the regions of interest from the background. In the literature, there are mainly two dual approaches. The contour segmentation approach which consists in locating the boundaries of objects and the region segmentation approach which partitions the image into a set of regions. Each region defines one or more related objects. Each of the two approaches of its nature has its advantages and disadvantages.
Some new attempts have been made in order to cooperate between these two segmentations by trying to combine the advantages of each one taken separately: the precision and the speed of a segmentation by contour, the closing of the borders and the density of the extracted information from segmentation by region \cite{Ref1}.

Compared to 2D segmentation, 3D segmentation gives a more complete understanding of a scene because 3D data (for example: multispectral images, point clouds, projected images, voxels and meshes) contains geometric information , richer in shape and scale with less background noise.

In this work, we propose to take advantage of a pioneering and recent architecture based on deep learning, used for the segmentation of 3D dental images. MeshSegNet consists of a complex Convolutional Neural Network (CNN), which automatically processes images of dental scans represented by triangular meshes, and which generates in output in a fairly precise way 15 membership classes which correspond to the different types of teeth as well as the gingiva \cite{Ref2}.

This paper is organized as follows:  the next section  presents a review of related works devoted to 3D dental image segmentation problem. Section~\ref{sec:3} provides a detailed description of our chosen algorithm. Section~\ref{sec:4} is consecrated to experimentations, results, some analysis and the comparison of different scenarios during training step. Finally, a critical discussion concludes the paper in Section~\ref{sec:5}.

\section{Related Work}
\label{sec:2}
In this section, we present some basic concepts related to dental anatomy and the subject of segmentation. Then, we provide a summary highlighting some done works in the field of dental segmentation based on Deep Learning.
\subsection{Segmentation, Anatomy and Dental Nomenclature}

\begin{figure}[H]
	\centering
	\includegraphics[width=.5\textwidth]{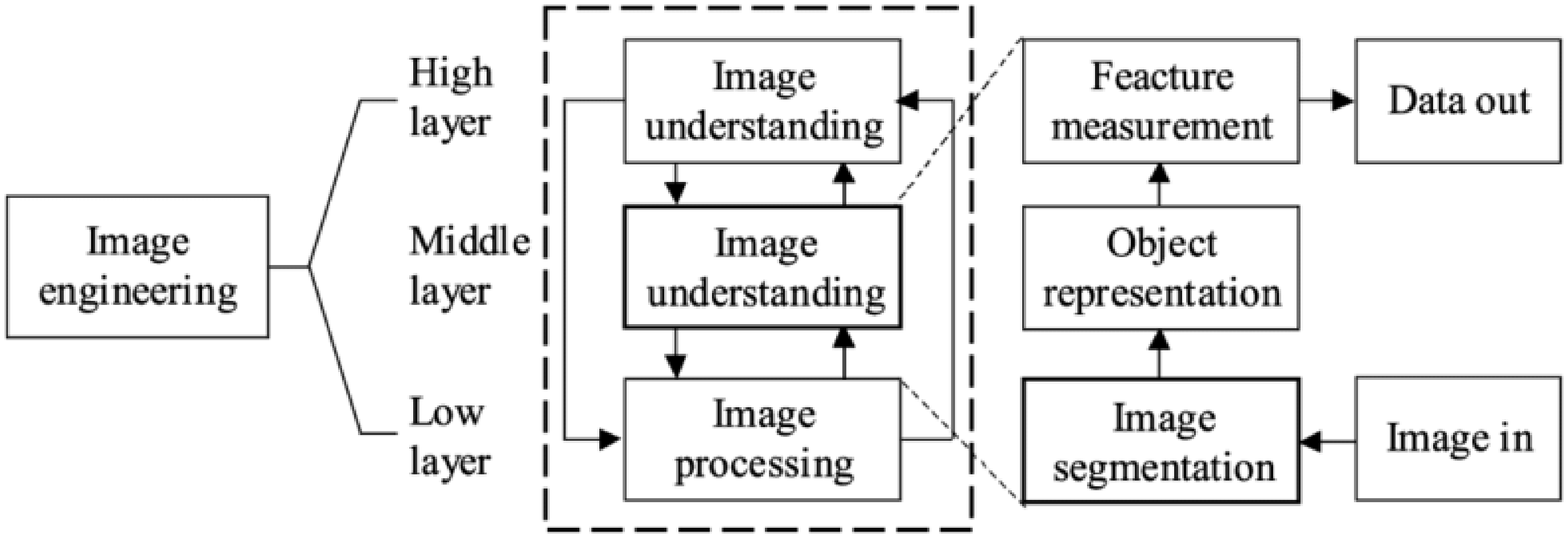}
	\caption{Different possible operations on the image \cite{Ref3}.}
	\label{fig:1}       % Give a unique label
\end{figure}

The image is one of the most privileged physical supports to transmit a message to our brain. It can also offer us the possibility of observing the majority of information received from the real world because it is part of many media that can be viewed by human beings, such as: multidimensional images, videos, animations, etc.
In order to better perceive images, understand their content and extract essential information, several techniques have been developed and many applications have been developed. According to \cite{Ref3}, all image operations can be grouped into three main layers, as shown in Fig. \ref{fig:1}.
As can be seen, image segmentation is the first step and also one of the most critical tasks in image analysis. It is undoubtedly the task that mobilizes the most effort in this layer.

Image segmentation can lead to the development of several interesting fields, strongly linked to image processing, such as: space exploration, medical research, logistics applications, vegetation studies, document processing, etc.

\begin{figure}[H]
	\centering
\begin{tabular}{c}
	\includegraphics[width=.5\textwidth]{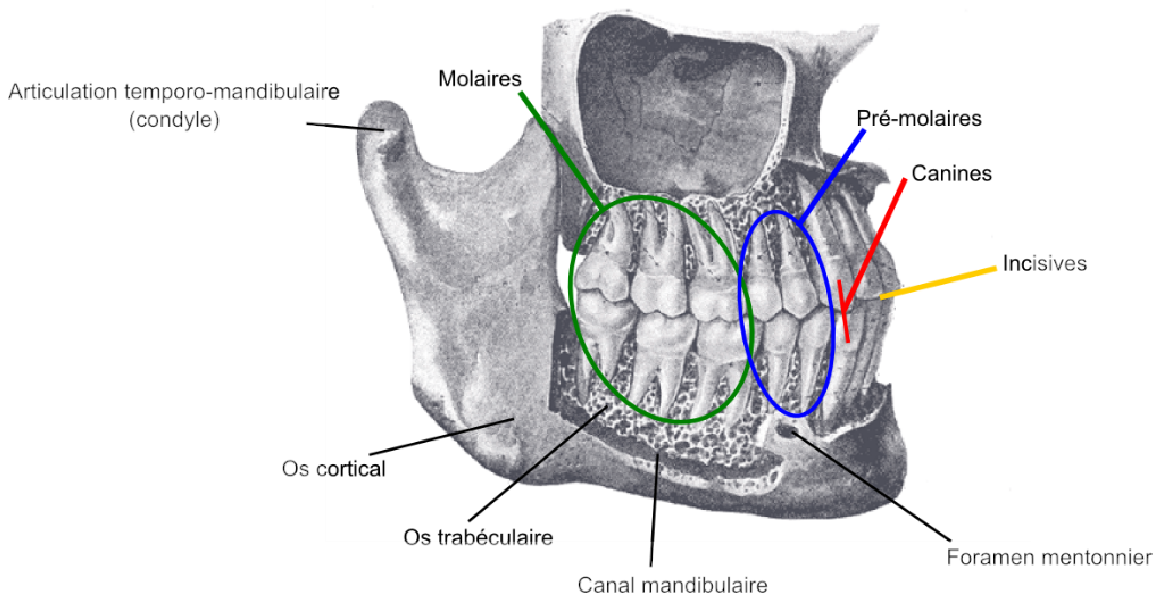} \\
	(a) \\	\includegraphics[width=.35\textwidth]{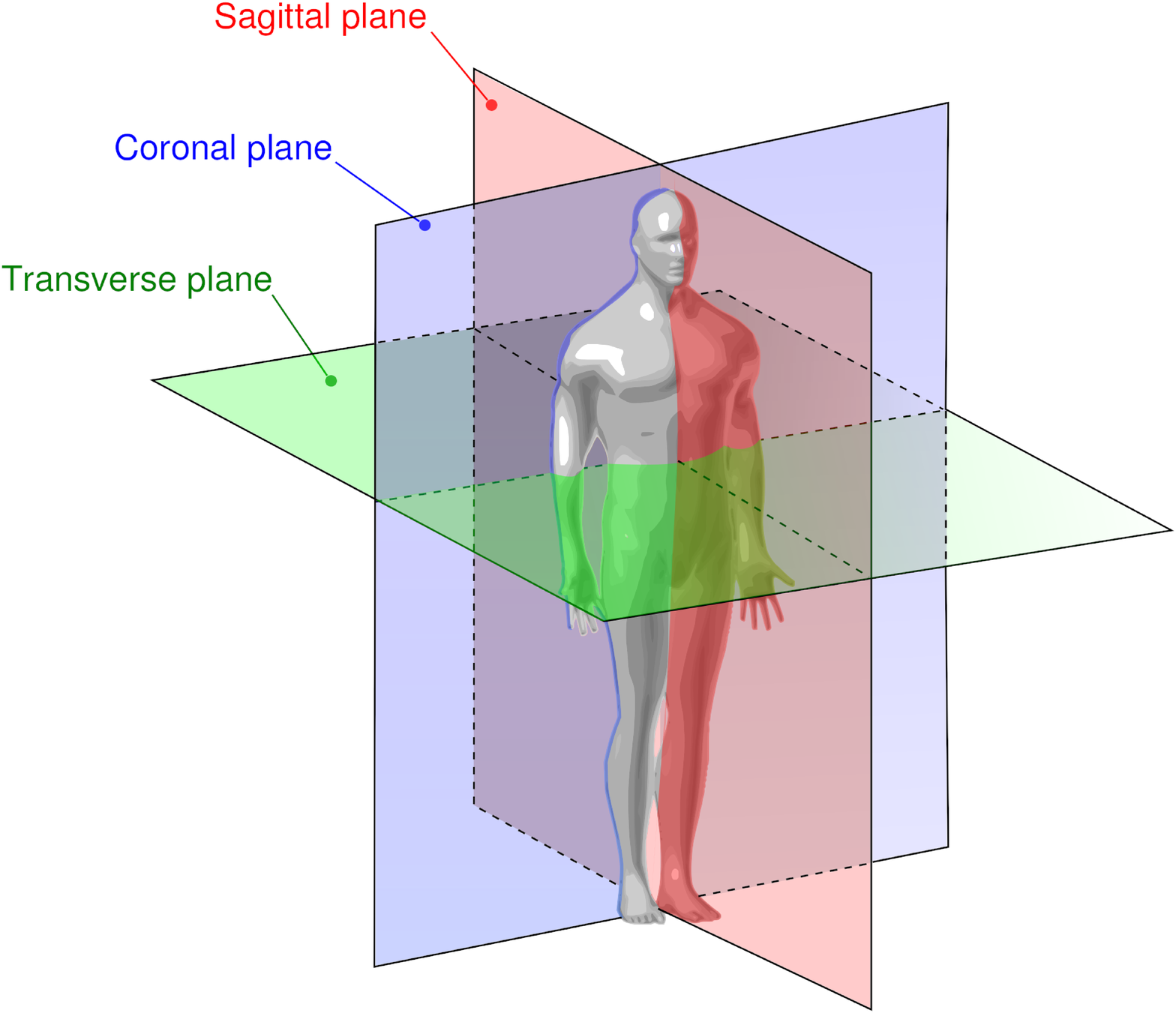}\\
	(b)\\
\end{tabular}
	\caption{Presentation of the dental anatomical context, (a) Structures of the maxillofacial zone, (b) Nomenclature of the anatomical planes \cite{Ref4}.}
	\label{fig:2}       % Give a unique label
\end{figure}

\begin{figure}[H]
	\centering
	\includegraphics[width=.5\textwidth]{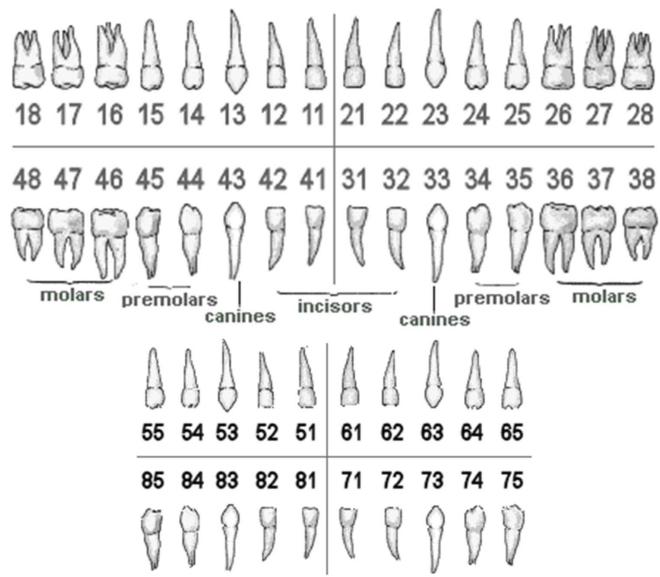}
	\caption{ISO notation system. Permanent teeth (between 11 and 48) and temporary teeth (between 51 and 85): top=maxilla, bottom=mandible, left=right side of the patient and right=left side of the patient \cite{Ref4}.}
	\label{fig:3}       % Give a unique label
\end{figure}

Teeth are the fundamental structures for the maxillofacial area. They constitute the first phase of the feeding mechanism which allows us to subsidize our vital needs. Depending on their type, they perform a very precise function that allows the first structure of the nutrient bowl to be reconstituted. The anterior teeth (incisors and canines) are used to cut the food into coarse pieces; which will be further reduced to finer pieces by the posterior teeth (premolars and molars). This biophysical transfiguration is associated with a chemical intervention which is ensured by the saliva and which constitutes the first stage for digestion and which then makes it possible to extract the essential nutrients for the survival of the human being (see Fig. \ref{fig:2}).
The dental environment is constantly evolving, which explains the particular attention paid to these crucial organs and why the majority of clinical practices are attached to them. In addition, these organs present a particular transformation comparing to the rest of the body in three cycles: first the establishment of the temporary teeth (they are also called "milk teeth", they paired between 6 months and 3 years), followed by a complete renewal by the permanent teeth (which generally takes place from 6 to 12 years), this ends with the springing of the third molars (namely the “wisdom teeth”, around 20 years) \cite{Ref4}. There are several nomenclatures to identify these teeth. The most widespread, and the one recommended by the World Health Organization, (i.e. the ISO system, see Fig. \ref{fig:3}).
Image segmentation methods have been improved over the past decades, but it remains a complex and difficult process due to the dissimilarity in these images. Much research has been done on the segmentation of dental x-rays. However, the analysis of dental images has some difficulty in comparing with other medical images.
Thus, the segmentation of dental images remains problematic because of:
\begin{itemize}
	\item A wide variety of topologies.
	\item The complexity of medical structures.
	\item Poor image quality caused by conditions such as low contrast and noise.
	\item Irregular and blurred contours.
	\item The presence of metal dental brackets used for alignment.
	\item Generation of artefacts during the acquisition process.
	\item Variation of tooth sizes.
	\item Space of missing teeth.
	\item Presence of superfluous teeth in somewhat rare cases; leading to unsuccessful segmentation.
\end{itemize}
Due to all these problems, it is always difficult to find an accurate and appropriate segmentation method for all types of dental X-ray images \cite{Ref5}.
\subsection{Previous Works}
In this part, we show some crucial works that have been done in the field of 3D image segmentation based on Convolutional Neural Networks:
\subsubsection{PointNet}
\begin{figure}[H]
	\centering
	\includegraphics[width=.5\textwidth]{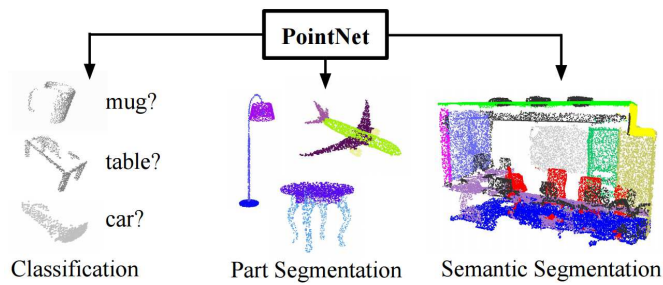}
	\caption{PointNet: Unified architecture for different tasks \cite{Ref9}.}
	\label{fig:4}       % Give a unique label
\end{figure}

\begin{figure}[H]
	\centering
	\includegraphics[width=.5\textwidth]{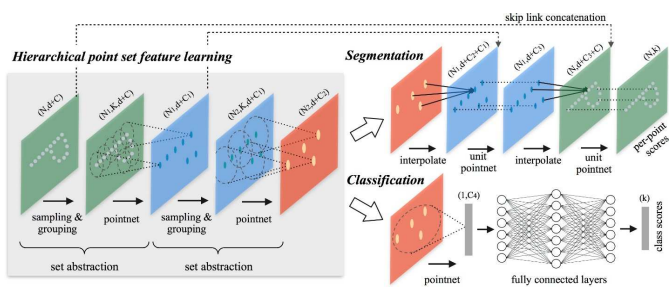}
	\caption{PointNet++: Hierarchical deep learning on point clouds in a metric space \cite{Ref10}.}
	\label{fig:5}       % Give a unique label
\end{figure}

Qi et al. proposed a new and pioneering deep network architecture that deals with point clouds (as unordered sets of points) (see Fig. \ref{fig:4}). The point cloud represents a common type of geometric data structure. Due to its irregular format, most researchers transform this data into grids (meshes) of regular 3D voxels or collections of images. However, this makes the data a bit large and causes some technical issues. In this article, the authors have designed a new type of neural network which directly manipulates point clouds, and which respects well the permutation invariance of the points in the input. This network, named PointNet \cite{Ref9}, provides a unified architecture for applications ranging from object classification to room segmentation and semantic scene analysis.
Other attempts were carried out by the creators of this CNN, thus giving rise to the PointNet++, the objective is to respect the spatial localities of these sets of points and to deal with the non-uniform densities in the natural point clouds \cite{Ref10}. PointNet++ incorporates hierarchical features with varying increasing scales. In addition, special layers capable of intelligently aggregating information at different scales have been implemented (see Fig. \ref{fig:5}).
Although it is simple and quite effective for the segmentation of primitive objects, PointNet does not correctly take into account the neighborhood relations between points which makes it of little use for the segmentation of 3D dental images. Moreover, the work of Lian et al. registred low mean values of DSC, SEN and PPV when used to segment 3D tooth surfaces (~78\%, 82\% and 76\% respectively).
\subsubsection{ToothNet}
\begin{figure}[H]
	\centering
	\includegraphics[width=.5\textwidth]{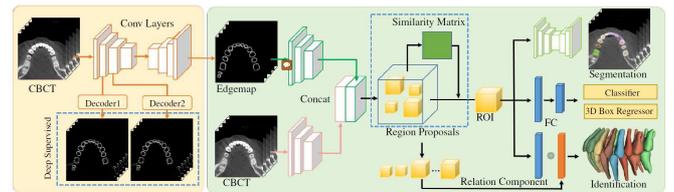}
	\caption{ToothNet: Two-Stage Network Architecture for Dental Image Segmentation and Identification \cite{Ref11}.}
	\label{fig:6}       % Give a unique label
\end{figure}

Cui et al. proposed a two-step deep neural network. In the first step, they extract the contour map of CBCT (for, Cone Beam Computed Tomography, or volumetric imaging by cone beam) images by a supervised deep network. In the second step, they concatenate the contour map features with the original image features and send them together to the 3D RPN (for, Region Propose Network 3D) network using a similarity matrix to filter out redundant inputs from the 3D RPN module. Four possible paths are then available, namely: segmentation, classification, 3D regressor and identification. During identification, they incorporate an additional component of spatial relationship between teeth to help disambiguate. Fig. \ref{fig:6} shows an overview of this solution \cite{Ref11}.
However, there are two cases of failure of this method:
Segmentation goes wrong in case of high grayscale value in the CT image (example of presence of metal artifact due to dental implants. Also, identification malfunctions if the tooth has wrong orientation. In addition, this network fails to correctly detect incomplete teeth.
\subsubsection{TSegNet}
\begin{figure}[H]
	\centering
	\includegraphics[width=.5\textwidth]{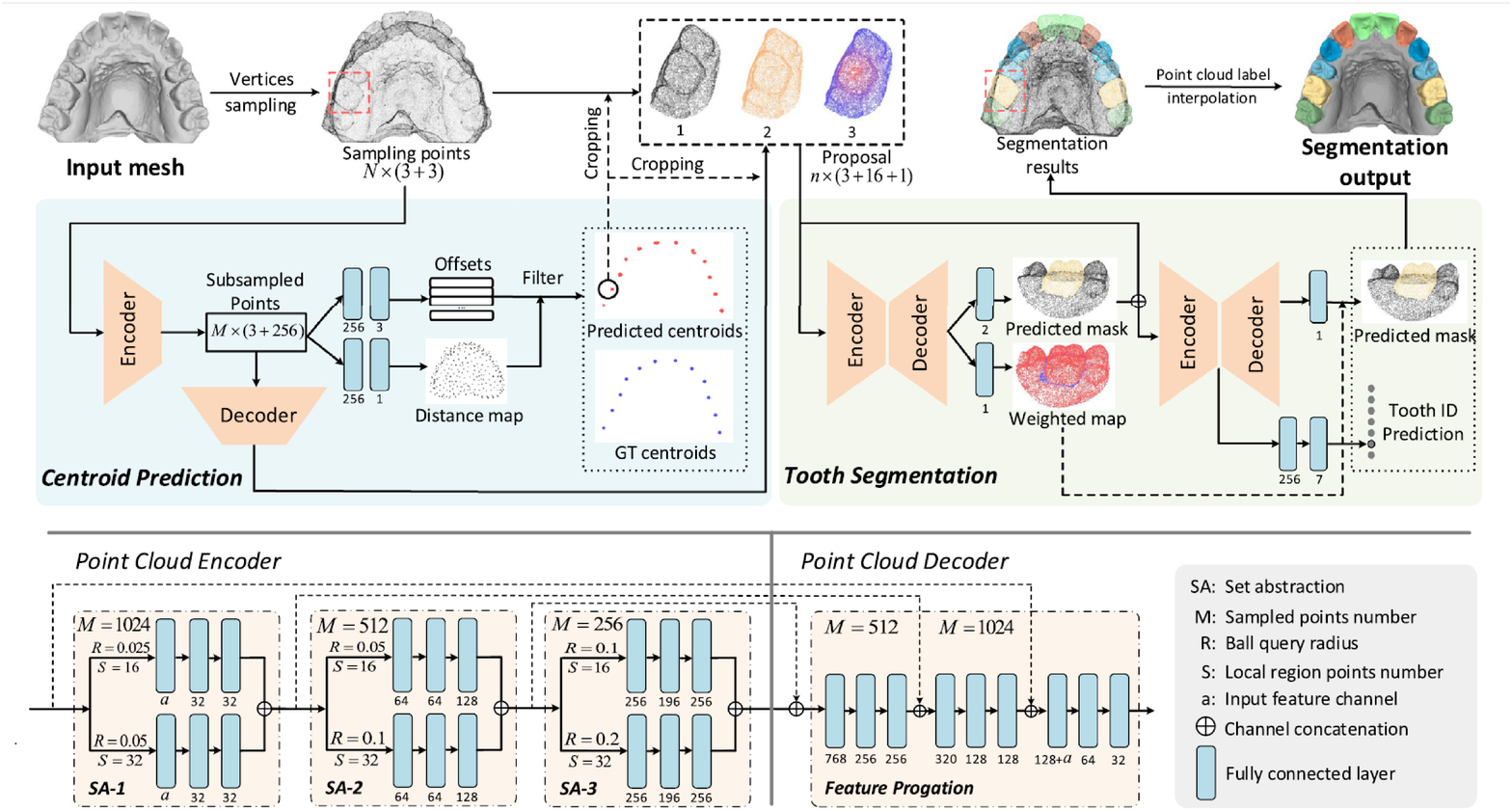}
	\caption{TSegNet: Two-Stage Network Architecture for Centroid Prediction and Dental Image Segmentation \cite{Ref12}.}
	\label{fig:7}       % Give a unique label
\end{figure}

Recently, Cui et al. proposed a novel end-to-end learning-based method, called TSegNet, for robust and efficient dental segmentation on 3D digitized point cloud data of dental models. This algorithm detects all teeth using a tooth centroid voting scheme, based on the distance in the first step, which ensures accurate location of dental objects even with irregular positions on abnormal dental models. Then, a confidence-based cascading segmentation module in the second stage is designed to segment each tooth individually, guided by the tooth's predicted centroid, and to resolve ambiguities caused by difficult cases: such as dental models with missing, crowded or misaligned teeth before orthodontic treatments. Fig. \ref{fig:7} shows a detailed diagram of this solution \cite{Ref12}.
Although the proposed method has achieved decent tooth segmentation results and outperforms some previous methods, it has some limitations that are worth considering. A typical example is that it tends to produce incomplete tooth segmentation in some cases like wisdom tooth and rudimentary tooth. A possible reason is that these cases are quite rare and weakly seen by the network during the training phase. Moreover, as is the case with the PoinNet network, TSegNet does not take into consideration the neighborhood relationships between adjacent voxels.
\section{Methodology and Technology}
\label{sec:3}
Our method is based on a modified and adapted version of the famous MeshSegNet network, joined with post-processing step which is used to correct and improve the obtained results.
\subsection{Motivations of the Technical Choice}
The evaluation of 3D dental segmentation algorithms based on deep learning constitutes an important problem, both for the choice of an algorithm and its parameterization for a user and for the comparison with the existing of a new algorithm. by a researcher or developer. Each method studied has its own advantages and intrinsic disadvantages.

In the previous section, we provided a general presentation of the main methods of image segmentation based on CNNs. Namely: PointNet, ToothNet and TSegNet, while highlighting the different phases of each system, its own attributes and the obtained results.

\begin{figure*}[h]
	\centering
	\includegraphics[width=\textwidth]{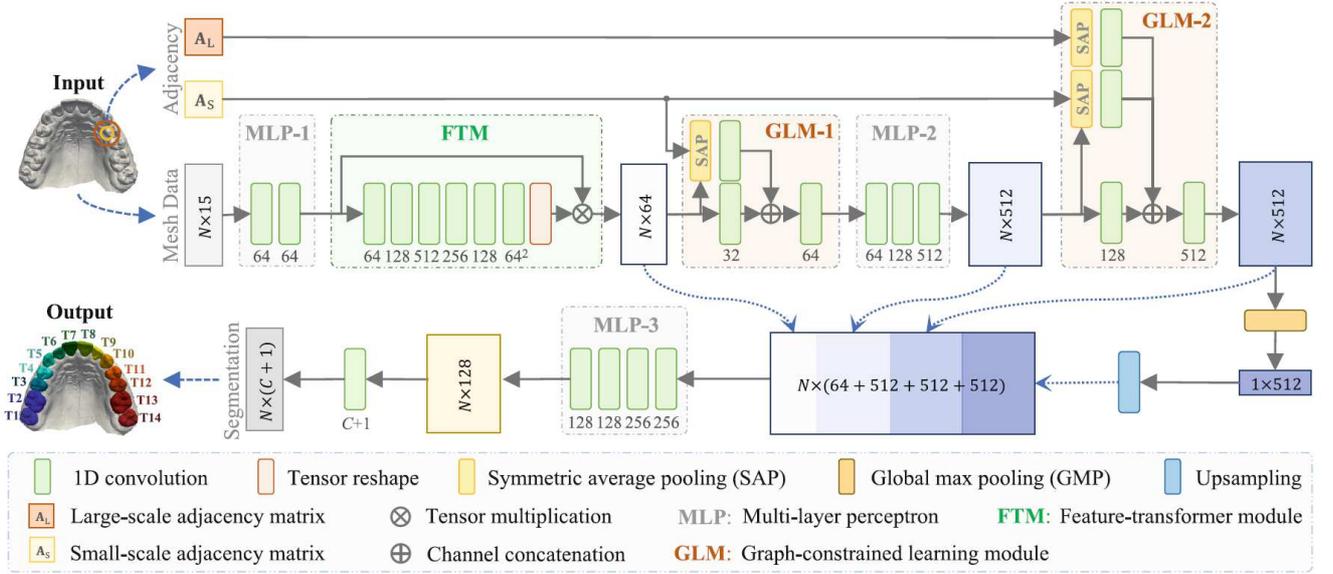}
	\caption{MeshSegNet: a multi-scale deep neural network that learns high-level geometric features used for end-to-end tooth segmentation on 3D tooth surfaces \cite{Ref13}.}
	\label{fig:8}       % Give a unique label
\end{figure*}

Our technical choice is finally focused on the MeshSegNet solution. Several strong points played in favor of it; which made it the most attractive and the most suitable for our problem, among others:

\begin{itemize}
	\item \textbf{PointNet:} Pioneer CNN in the field of 3D (segmentation and classification); Generic network (not specialized; no consideration of geometric links between points; source code available online, low accuracy value in 3D dental image segmentation.
	\item \textbf{ToothNet:} Specialized model; inadequate model: based on 2D images (CBCT); incomplete dental segmentation (wisdom teeth and rudimentary teeth); source code not available.	
	\item \textbf{TSegNet:} Input image on 3D cloud point; not robust to inclination and the high contrast of the GSV input image; source code not available.
	\item \textbf{MeshSegNet:} Best specialized architecture developed, source code available. A diagram of the selected approach is presented in Fig.~\ref{fig:8}.
\end{itemize}
\subsection{Detailed Steps}
Here, the authors proposed an end-to-end deep neural network to directly learn high-level geometric features from the raw mesh for automatic tooth segmentation. Specifically, the MeshSNet method extends the PointNet architecture in three aspects: (1) Points are replaced by meshes, as the meshes link them topologically to show the local structure clearly; (2) They also provide graph-constrained multi-scale learning modules to explicitly take into account the local geometric context and mimic the learning of hierarchical features of CNNs; and (3) they densely merge cellular features, multiscale contextual features, and translationally invariant holistic features for cell annotation. Fig.~\ref{fig:8} shows an illustration of this architecture \cite{Ref13}. Here, $A_L$ and $A_S$ correspond to the square adjacency matrices (of size N×N) for the large and small radius, respectively calculated based on the Euclidean distance between the cell centroids (triangles or meshes).

\begin{figure}[H]
	\centering
	\includegraphics[width=.5\textwidth]{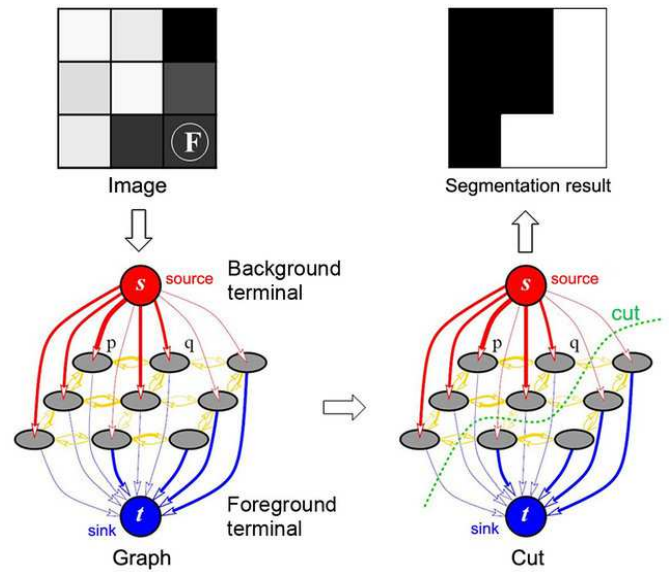}
	\caption{Graph-Cut: Example of Graph-Cut segmentation for a simple 3×3 image \cite{Ref15}. Its principle remains valid for 3D images (in voxels).}
	\label{fig:9}       % Give a unique label
\end{figure}

The segmentations produced by this deep neural network (MeshSegNet) can lead to isolated false predictions or unsmooth edges. A graph-based post-processing (e.g. Graph-cuts \cite{Ref14,Ref15}) attempting to minimize a certain energy function E, is thus applied for the refinement of the results produced in both image segmentation tasks and 3D tooth surface labeling. An example is shown in Fig.~\ref{fig:9}.

\section{Experimental Results}
\label{sec:4}
We begin this section with a brief introduction relating to the performance measures that we used for the evaluation and then reveal the test bases. The training protocols are then discussed. After that, numerical results and some relevant comparisons are given. We end by providing a brief run-time study of our method and the involved devices.
\subsection{Evaluation Criteria}
\label{sec:41}
For training, four measures are selected: Loss (using the CCE function(y, \^{y})) (Categorical Cross Entropy), DSC (Dice Similarity Coefficient), the Sensitivity (or the Selectivity ) (SEN) and the Positive Predictive Value (PPV). However, the Mean Average Precision (MAP) is used for the evaluation of this approach on a subset of randomly selected data.
\subsubsection{DSC}
It represents a statistical indicator that measures the similarity of two samples. This index measures the presence or absence of entities in these sets:

\begin{equation}
	\mathit{DSC} = \frac{2 \times TP}{(2TP + FP + FN)} 
\end{equation}

\subsubsection{SEN}

In medicine, the sensitivity of a diagnostic test is its ability to detect a maximum of positive results when a hypothesis is verified (i.e. to have the fewest false negatives):
\begin{equation}
	\mathit{SEN} = \frac{TP}{(TP+FN)} 
\end{equation}
\subsubsection{PPV}

The predictive value is the probability that a condition will be met based on the result obtained. The test must be binary, i.e. it can only give two different values:

\begin{equation}
	\mathit{PPV} = \frac{TP}{(TP+FP)} 
\end{equation}

\subsubsection{MAP}

For a set of examples, the MAP designates the average score of the average precision values, associated with each of them:

\begin{equation}
	\textit{MAP} = \frac{1}{n} \sum\limits_{i=1}^{n} \textit{AP}_i = \frac{1}{n} \sum\limits_{i=1}^{n} \frac{\textit{TP}_i}{(\textit{TP}_i+\textit{FP}_i)}
	\label{eq:10}
\end{equation}
\begin{figure}[H]
	\centering
	\includegraphics[width=.5\textwidth]{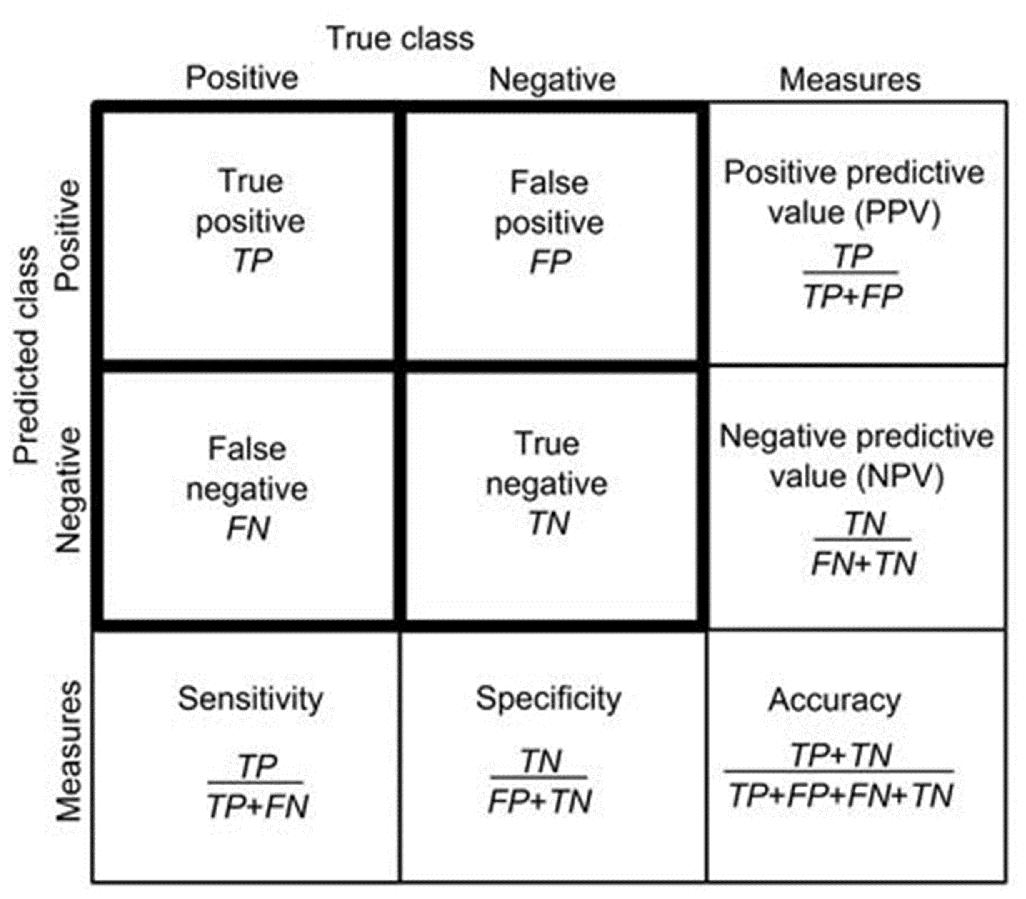}
	\caption{Contingency matrix and calculated measures based on the results of predictions from the training set.}
	\label{fig:10}       % Give a unique label
\end{figure}

\subsection{Datasets}
\label{sec:42}
To test this segmentation approach (MeshSegNet), we applied it to a data sample of 95 3D images, from 2 different databases:
\begin{enumerate}
	\item Benchmark of 85 supervised synthetic 3D images (annotated or labeled) taken of the dental maxilla in .obj format used during the training phase (including 10 dedicated to the test phase).
	\item Benchmark of 10 unsupervised (non-segmented) 3D synthetic images taken of the format dental mandible. stl which are thus all devoted to the test phase.
\end{enumerate}

All these private data were provided by the CORUO company, and in order to adapt its real class values to the CNN MeshSegNet (accepting only 15 classes of teeth including the gingiva), we applied a Clipping operation (all classes to the above 14 will automatically be assigned to a single class (wisdom tooth)).

\subsection{Training Set and Parameters Adjustement}
\label{sec:43}
In order to increase the training data set, we augment the available 3D dental scans (more precisely their meshes) by applying random geometric transformations (within reasonable ranges) of type (1) Rotation (along the three axes XYZ) in a closed interval of [-15°, +15°], (2) Scale change: [0.8 , 1.2] and/or (3) Translation (also along the three XYZ axes): [-10 , +10]). This generated a total of 85 × 20 = 1700 additional examples.

In this work, Intraoral scans are stored in VTP (VTK-like polygonal data) format. To read, write and manipulate VTP files, we use the famous Python-library Vedo.

This new base will be injected into the initial model to ensure continuous training taking into account its latest Pytorch backup (including its parameters, its synaptic weights of neurons/biases, etc.) provided by the authors of MeshSegNet networks (960 pre-trained examples ) (or completely separate learning (reset) with or without data boost), resulting in 4 different learning patterns.

Here, we have retained the same values proposed by the builders of the MeshSegNet model to set the hyper-parameters. Thus, we train it using ADAM with a Batch size of 10 (it characterizes the number of words to be injected into the network before updating its weights, and averages over a number of images of 10 , 100, or 1000. Smaller batch values have been shown to improve generalization performance). This value will be reduced to 2 (minimum allowed) when it comes to training the raw database without resorting to data augmentation (reason: very few examples, objective: to optimize training metrics). 

The set of samples is trained for 200 epochs (Epochs: the number of training steps performed on the complete training set, which allows the convergence of the ADAM optimizer). The weight matrix $\omega_j$ and the bias vector $b_j$ are continuously updated throughout the learning phases. 

The hyperparameters (which control the weights and the biases) are set using the values recommended by the authors: in our tests, we set the learning rate $\gamma$ to $10^{-4}$, the coefficients used to calculate the moving averages of the gradient and its square ($\beta$1= 0.9, $\beta$2=0.999). The stability term EPS= $1e^{-8}$ \cite{Ref6,Ref7,Ref8}.

Finally, the weights decay parameter (Decay, which represents another term in the weights update rule and is used to exponentially decrease the weights down to zero, in case no further updates are expected) is conventionally fixed at $\epsilon$ (infinitely small ~0).
However, regularization (Dropout) was applied during training, which helps our model to generalize better and obtain higher accuracy and to prevent the phenomenon of Overfitting. By randomly deactivating neurons at each stage of forward/backward propagation. The rate applied is equal to 0.5.

From the triangular meshes provided and the 15 assignment classes, the model automatically deduces their normal vectors as well as their relative positions with respect to the entire surface of the object to form the 15 input attributes of the model. It also performs 3D image normalization to ensure that each input parameter (voxel, in this case) has a similar data distribution. This keeps all data at the same scale, preserves relative information, and speeds up convergence. Finally, some important Python libraries are imported (such as: Numpy, Pytorch with GPU option, Pandas, OS, Datetime, Sickit-learn, Visdom, Vedo, Pygco and many others).

\subsection{Results}
\label{sec:44}
Table~\ref{tab:1} presents the final obtained results concerning the different tests carried out on the first data set during training. We can clearly observe that the registred values of the learning performance indices without data augmentation are remarkably poorer than those with data augmentation (this can be reflected by a high rate of Loss and low values of other metrics) .
\begin{table*}[h]
	\centering
	\caption{Training output measurement values (at the end of training) [\%].}
	\label{tab:1}       % Give a unique label
	\begin{tabular}{lllll}
		\hline\noalign{\smallskip}
		\multirow{2}{*}{Metrics} &  \multicolumn{2}{c}{Without Data Augmentation}&  \multicolumn{2}{c}{With Data Augmentation} \\ 
		& Isolated Training & Continuous Training & Isolated Training & Continuous Training \\
		\noalign{\smallskip}\hline\noalign{\smallskip}
		Loss & 89.82 & 85.79 & \textbf{01.99} & 12.69 \\ 
		
		DSC & 10.23 & 14.24 &\textbf{98.04} & 87.32 \\ 
		
		SEN & 24.39 & 23.77 &\textbf{97.99}& 92.27 \\ 
		
		PPV & 30.10 & 34.74 & \textbf{98.19} & 90.52 \\ 	
		\noalign{\smallskip}\hline
	\end{tabular}
\end{table*}
This can be justified by the phenomenon of Underfitting, in which this model remains unable to provide accurate predictions. Therefore, the cost of error in the learning phase remains high. In fact, in this case (85 images only) we have very little data to build our model perfectly and also these data are heterogeneous.

\begin{figure}[H]
	\centering
	\begin{tabular}{c}
		\includegraphics[width=.4\textwidth]{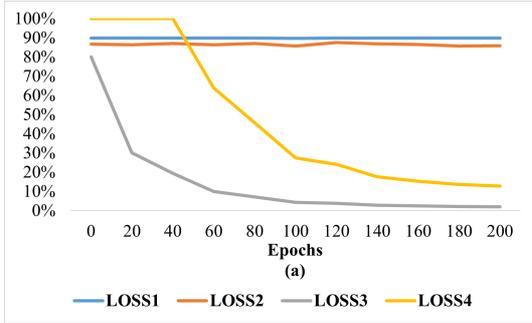} \\
		(a) \\	\includegraphics[width=.4\textwidth]{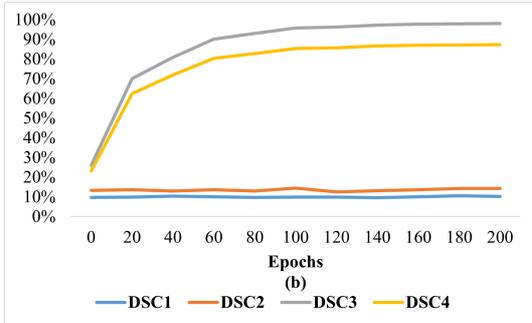}\\
		(b)\\
		\includegraphics[width=.4\textwidth]{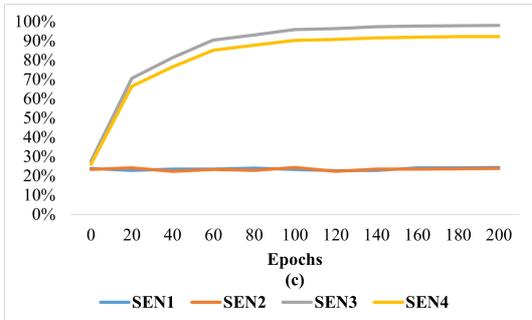} \\
		(c) \\	\includegraphics[width=.4\textwidth]{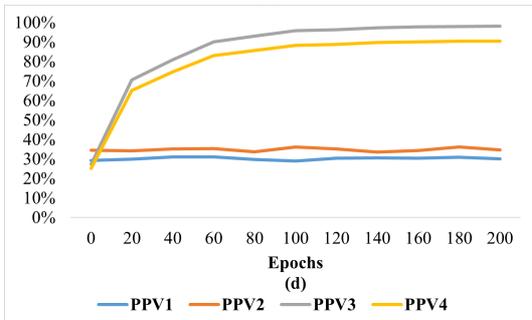}\\
		(d)\\
	\end{tabular}
	\caption{Evolution of metrics during different training epochs during 200 epochs, (a) Loss, (b) DSC, (c) SEN, (d) PPV. 1: Isolated training without data augmentation, 2: Continuous training without data augmentation, 3: Isolated training with data augmentation, 4: Continuous training with data augmentation.}
	\label{fig:11}       % Give a unique label
\end{figure}

For more details, Fig. \ref{fig:11} shows the variation of the above mentioned measurements over the whole training phase (for 200 epochs).
Similarly, and similarly to what we obtained during the training phase, Table \ref{tab:2} presents a detailed numerical comparison of the four approaches in terms of MAP during the test phase.

The separate learning approach with data augmentation was able to outperform all other approaches. On the other hand, it remains prohibitive in terms of execution time necessary for the completion of the training operation.
\begin{table*}[h]
	\centering
	\caption{Comparison of the different approaches in terms of execution time required during the training phase [HH:MM:SS] and in map [\%] during the test phase.}
	\label{tab:2}       % Give a unique label
	\begin{tabular}{lllll}
	\hline\noalign{\smallskip}
	\textbf{Approach}	& Continuous learning without DA & Continuous learning with DA & Continuous learning without DA & Continuous learning with DA \\
	\noalign{\smallskip}\hline\noalign{\smallskip}
	\textbf{Time (Training) }& 00:37:37&	19:18:02&	00:37:04&	15:28:19 \\ 
	
	\textbf{MAP (Test)} & 40.00\%	&81.60\%&	31.20\%&	84.00\% \\ 	
	\noalign{\smallskip}\hline
\end{tabular}
\end{table*}

\begin{figure*}[h]
	\centering
	\begin{tabular}{ccc}
		\includegraphics[width=.3\textwidth]{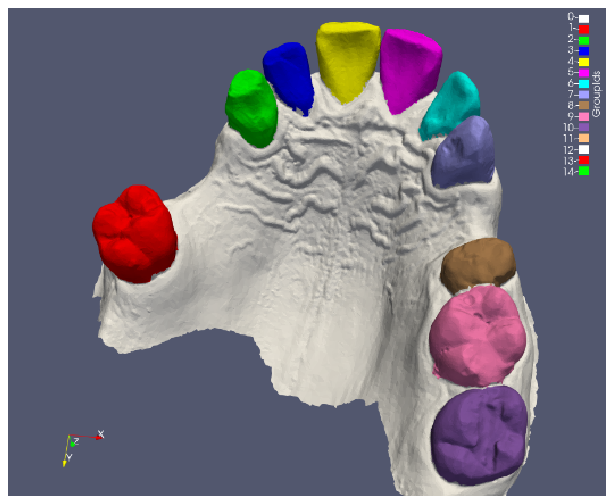} & \includegraphics[width=.3\textwidth]{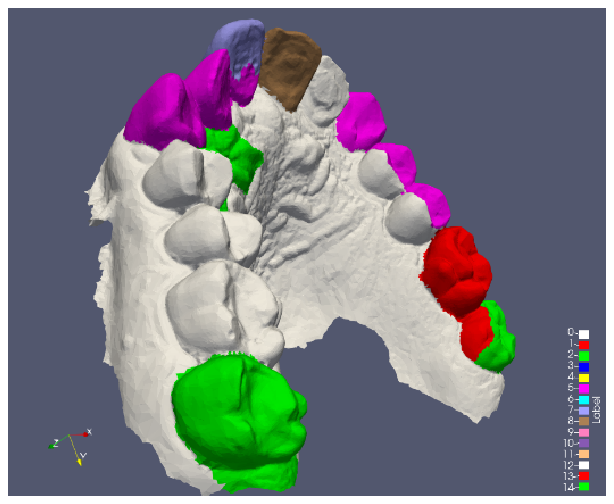} & \includegraphics[width=.3\textwidth]{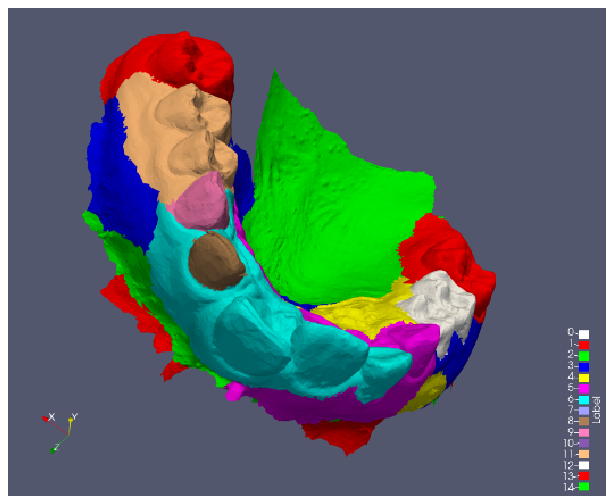} \\
		(a) & (b) & (c) \\
		\includegraphics[width=.3\textwidth]{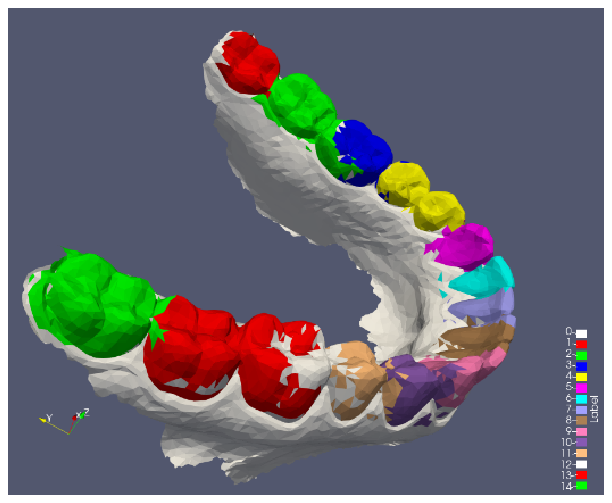} & \includegraphics[width=.3\textwidth]{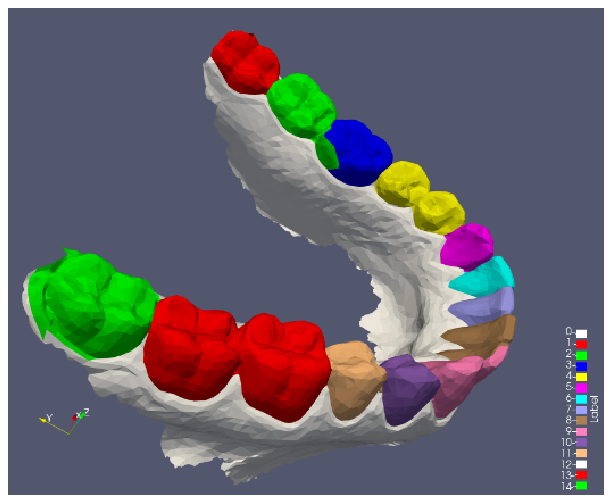} & \includegraphics[width=.3\textwidth]{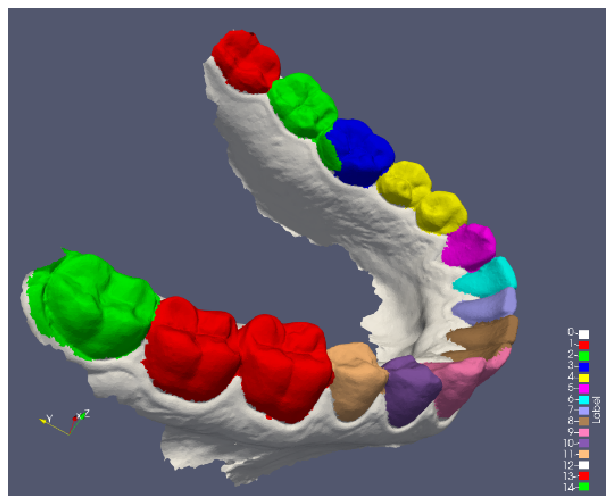} \\
		(d) & (e) & (f) \\
	\end{tabular}
	\caption{Qualitative analysis of dental segmentation results using deep learning based approach (MeshSegNet), (a) Accurate segmentation, (b) Incomplete segmentation, (c) Poor segmentation outcome, (d) Applied segmentation prediction on the compressed image (rate = 10\%), (e) Post-processing (graphic cut method), (f) Restoration (decompression ×10). Line 1: Examples taken from maxilla scans, Line 2: Examples taken from mandible scans.}
	\label{fig:12}       % Give a unique label
\end{figure*}

However, the obtained value of MAP increases to 80.15\% when we test the MeshSegNet model on the mandible test basis (Benchmark 2). In summary, the test goes through the following steps:
\begin{enumerate}
	\item Compress the input image (downsampling): using the Decimate(rate\%) method.
	\item Predict the result (of segmentation) using the MeshSegNet architecture applied to the compressed image.
	\item Post-processing: through the graph-cuts method used to refine the predicted result by smoothing tooth contours and eliminating isolated false predictions (get homogenizing regions).
	\item Decompress the corrected image (over-sampling) using the KNN method (Number of neighbors = 3), ceiling = 100k cells (triangular meshes), based on their barycenters.
\end{enumerate}
A detailed and explanatory qualitative visual comparison concerning the segmentation images provided by this technique applied to maxillae and a mandible is given in Fig. \ref{fig:12}.
\subsection{Computational Time}
\label{sec:45}
The various experiments were carried out at the Computing Center of the University of Burgundy (CCUB). It is made available to researchers, offering an intensive computing and data storage platform for the parallelization, optimization and debugging of computer codes. Each workgroup is entitled to an assembly of 32 processors (Intel(R) Xeon(R) CPU E5-2667 v4 at 3.20 GHz, 25 MB L3 cache, 8 cores and 2 threads per core) and 256 GB of RAM memory used by the CentOS distribution. The high-end Tesla K80 GPU is also included (with 2×12 GB of GDDR5, Number of floating point operations per second = 5.59 Tflops, Bandwidth = 447 GiB/s, of which a total of 4992 compute units )).
  
\begin{figure}[H]
	\centering
	\includegraphics[width=.48\textwidth]{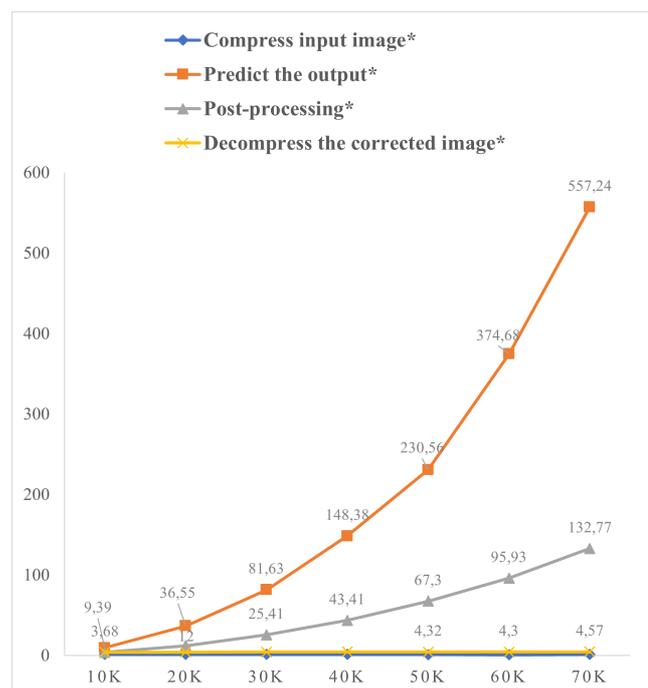}
	\caption{Evolution of average execution time per image (in seconds) according to different sizes of 3D images of mandibles during the test phase.}
	\label{fig:13}       % Give a unique label
\end{figure}

\begin{table*}[h]
	\centering
	\caption{Summary of the average calculation time per image (in seconds) during the different phases of MeshSegNet applied to a dataset of 10 images of mandibles in the ".stl" format with different compression rates.}
	\label{tab:3}       % Give a unique label
	\begin{tabular}{llllllll}
		\hline\noalign{\smallskip}
		\textbf{Average execution time per image (µ)}	& \textbf{10k} & \textbf{20k} & \textbf{30k}& \textbf{40k}& \textbf{50k} & \textbf{60k}& \textbf{70k} \\
		\noalign{\smallskip}\hline\noalign{\smallskip}
		Compress input image & 1.21&	1.14&	1.18&	1.08&	1.04&	1.01&	1.13 \\ 
		
		Predict result & 9.39&	36.55&	81.63&	148.38&	230.56&	374.68&	557.24 \\ 	
		Post-processing & 3.68&	12.00&	25.41&	43.41&	67.30&	95.93&	132.77 \\ 
		Decompress corrected image & 4.11&	4.08&	4.28&	4.39&	4.32&	4.30&	4.57 \\ 
		\noalign{\smallskip}\hline
	\end{tabular}
\end{table*}
The registred running time (in seconds) of the different phases by varying the input 3D image compression ratio (downsampling) when using the MeshSegNet architecture is shown in Table \ref{tab:3}. We can indeed realize that the calculation times were reasonable with regard to the size of the 3D images. However, time exploded exponentially, especially for the Prediction and Post-processing phases (as shown in Fig. \ref{fig:13}).
\section{Conclusion and Discussions}
\label{sec:5}
In this paper, our main contribution is the implementation of an existing and efficient method, that is used to segment 3D teeth images based on Deep Learning. For that purpose, a modified and adapted version of the famous MeshSegNet network is proposed.

MeshSegNet represents an interesting solution because it avoids the possible limitations of previous solutions (i.e. the non-consideration of neighborhood relations between voxels of the original PointNet 3D network), it improves the classification process and it does not need special Pre-processing to make it work. In addition, it automatically incorporates a Post-processing procedure to correct the primary generated results.

The evaluation indices provided by this approach, whether numerical or visual, have shown its effectiveness and its robustness to variations (rotations, change of scale, etc.). But, like all research and development methods, the latter has registered some drawbacks during execution such as:

\begin{itemize}
	\item The number of parameters, which is relatively big, the results strongly depend on the optimization of these parameters.
	\item The calculation time is sometimes very long, estimated at a few hours due to several endogenous and exogenous factors.
\end{itemize}

Many perspectives can be cited to improve this work and enrich our study. Among them, we can mention:

\begin{itemize}
	\item Make improvements to the detection procedure so that each tooth will be extracted separately.
	\item Consider an additional procedure that effectively addresses the overlapping of two close teeth upon detection.
	\item Propose a segmentation of 3D images by mutual cooperation (Region/Edge).
	\item Find an automatic adaptation of the CNN network parameters to the characteristics of the input image.
\end{itemize}

\section*{Acknowledgments}
This research was supported by XLIM Laboratory and Coruo Entreprise who provided insight and expertise that greatly assisted the research.

\bibliographystyle{unsrt}
\bibliography{bibliography}

%\begin{thebibliography}{99}
%
%\bibitem{ange96} A. Bonnaccorsi. On the Relationship between Firm
%Size and Export Intensity, \emph{Journal of International Business
%Studies}, XXIII (4), pp. 605-635, 1992. (journal style)
%
%\bibitem{caves96} R. Caves. M\emph{ultinational Enterprise and
%Economic Analysis}, Cambridge University Press, Cambridge, 1982.
%(book style)
%
%\bibitem{clerc99} M. Clerc. The Swarm and the Queen: Towards a
%Deterministic and Adaptive Particle Swarm Optimization. In
%\emph{Proceedings of the IEEE Congress on Evolutionary Computation
%(CEC)}, pp. 1951-1957, 1999. (conference style)
%
%\bibitem{crokell86} H.H. Crokell. Specialization and International
%Competitiveness, in \emph{Managing the Multinational Subsidiary}, H.
%Etemad and L. S, Sulude (eds.), Croom-Helm, London, 1986. (book
%chapter style)
%
%\bibitem{deb2000} K. Deb, S. Agrawal, A. Pratab, T. Meyarivan. A
%Fast Elitist Non-dominated Sorting Genetic Algorithms for
%Multiobjective Optimization: NSGA II. \emph{KanGAL report 200001},
%Indian Institute of Technology, Kanpur, India, 2000. (technical
%report style)
%
%\end{thebibliography}

\vspace{5mm}

%\noindent{\bf\Large Author Biography} \vspace{5mm}

%{\bf Omar BOUDRAA} is a Doctor on Distributed and Mobile Computing, a Temporary Research Assistant and Assistant Lecturer at Computer Science Department, University of Limoges, Limoges, France. Also, he was a visiting researcher at LIB Laboratory of Burgundy University, Dijon, France. His teaching and research interests are in image processing, artificial intelligence, networks and systems administration.

\end{document}